\documentclass[prb,twocolumn,superscriptaddress,longbibliography,notitlepage]{revtex4-1}
\usepackage{amssymb}
\usepackage{amsmath}
\usepackage{graphicx}
\usepackage{hyperref}
\usepackage{xcolor}
\usepackage{multirow}

\newcommand{\comment}[1]{}

\newcommand{\bra}{\langle}
\newcommand{\ket}{\rangle}

\begin{document}

\title{Solving Fermi-Hubbard-type Models by Tensor Representations of Backflow Corrections}

\author{Yu-Tong Zhou}
\affiliation{CAS Key Laboratory of Quantum Information, University of Science and Technology of China, Hefei 230026, China}
\affiliation{Synergetic Innovation Center of Quantum Information and Quantum Physics,University of Science and Technology of China, Hefei, 230026, China}
\affiliation{Hefei National Laboratory, University of Science and Technology of China, Hefei 230088, China}

\author{Zheng-Wei Zhou}
\email{zwzhou@ustc.edu.cn}
\affiliation{CAS Key Laboratory of Quantum Information, University of Science and Technology of China, Hefei 230026, China}
\affiliation{Synergetic Innovation Center of Quantum Information and Quantum Physics,University of Science and Technology of China, Hefei, 230026, China}
\affiliation{Hefei National Laboratory, University of Science and Technology of China, Hefei 230088, China}

\author{Xiao Liang}
\email{liangstein@psc.edu}
\affiliation{Pittsburgh Supercomputing Center, Carnegie Mellon University, Pittsburgh, PA 15213, USA}
\affiliation{Department of Physics, Carnegie Mellon University, Pittsburgh, PA 15213, USA}

\begin{abstract}

	The quantum many-body problem is an important topic in condensed matter physics. To efficiently solve the problem, several methods have been developped to improve the representation ability of wave-functions.
	For the Fermi-Hubbard model under periodic boundary conditions, current state-of-the-art methods are neural network backflows and the hidden fermion Slater determinant.
	The backflow correction is an efficient way to improve the Slater determinant of free-particles.
	In this work we propose a tensor representation of the backflow corrected wave-function, we show that for the spinless $t$-$V$ model, the energy precision is competitive or even lower than current state-of-the-art fermionic tensor network methods.
	For models with spin, we further improve the representation ability by considering backflows on fictitious particles with different spins, thus naturally introducing non-zero backflow corrections when the orbital and the particle have opposite spins.
	We benchmark our method on molecules under STO-3G basis and the Fermi-Hubbard model with periodic and cylindrical boudary conditions.
	We show that the tensor representation of backflow corrections achieves competitive or even lower energy results than current state-of-the-art neural network methods.

%Quantum many-body problem is an important topic in condensed matter physics. To efficiently solve the problem, several methods have been developped. Until now, a powerful structure of tensor network named as projected entangled pair states (PEPS) are promising. However, the application of PEPS is limited by its high contraction complexity, especially for systems with PBC (PBC).

\end{abstract}
\maketitle

\section{Introduction}
%1, importance of quantum many-body problem, traditional numerical methods have limitations
%2, neural network methods have achieved state-of-the-art, for psin models..., for fermi models, second q to spin model, or neural network corrections for wave-function. insight is correlation, correlation can represent ground energy
%3, backflow corrections on wave-functions, history and the new NN method.
%4, start from Green's function, which includes correlation.

Exotic physical phenomena emerge when a large number of microscopic particles interact with each other. Understanding phenomena such as superconductivity, quantum spin-liquid and the quantum Hall effect requires solving the quantum many-body problem to a high accuracy. However, solving the problem is challenging because the Hilbert space of the solution grows exponentially with respect to the size of the problem.

Several methods have been developped but there are still limitations. For example, exact-diagonalization (ED) has high accuracy but the problem size is limited~\cite{ED}. The density-matrix-renormalization-group (DMRG) can solve one-dimensional or quasi one-dimensional systems~\cite{DMRG}, but the accuracy is not satisfactory for two-dimensional systems.
Quantum-Monte-Carlo (QMC) has no limitation on dimensions and has high precision, but the computational complexity is too high for systems with the ``sign problem"~\cite{QMC}. The projected-entangled-pair-state (PEPS) can solve the two-dimensional system under open boudary condition (OBC) with a high accuracy, however the computational complexity is high especially for periodic boundary condition (PBC)~\cite{PEPS_review,fPEPS,PEPS_liu}.
Recently, neural networks (NN) have shown potential in representing quantum many-body states~\cite{RBM_science,RBM_tomography,CNN_J1J2_1,CNN_J1J2_2,CNN_J1J2_3,CNN_J1J2_4,CNN_J1J2_5,RBM_J1J2,NN_molecules,NN_solids_2,DNN schrodinger 1,NN backflow,DNN schrodinger 2,NN_fermi_tV,HFSD,autoregressive2,autoregressive3}.
%For example, the restricted-Boltzmann-machine (RBM) can solve the frustration free spin-1/2 Heisenberg model with high precision~\cite{RBM_science}. By considering the sign structure and the amplitude separately and employing a deep NN to represent the amplitude, the challenging frustrated spin-1/2 $J1$-$J2$ model under PBC has been solved to state-of-the-art precision~\cite{CNN_J1J2_5}.
For solving Fermi-Hubbard-type models, one approach is the Jordan-Wigner transformation on the Hamiltonian that treats the problem as solving a spin model~\cite{NN_molecules,NN_solids_2}. Another way is improving the single-particle Slater Determinants(SD) by NN backflow transformations~\cite{DNN schrodinger 1,DNN schrodinger 2,NN backflow} or multiplying the SD by a NN Jastrow factor~\cite{NN_fermi_tV}.
The state-of-the-art wave-function named the hidden fermion Slater determinant (HFSD) considers hidden fermionic particles and calculates the determinant of an enlarged matrix~\cite{HFSD}.

Solving the ground state of the Fermi-Hubbard model near 1/8 doping is important for understanding the mechanism of superconductiviey, however the ground state is challenging to solve~\cite{Hubbard_1,Hubbard_2,AFQMC_1,AFQMC_2,AFQMC_3}.
%The Hamiltonian of a Fermi-Hubbard-type model consists of hopping terms and interaction terms, and the ground state energy can be solved by one-body and two-body correlations without wave-functions\cite{energy_1}.
%However the wave-function is important for understanding the property of a quantum many-body state.
In mean-field theory, the Hamiltonian is in a quadratic form, and the ground state is the Hartree-Fock (HF) state.
The wave-function of the HF is a Slater determinant, which is an exact representation for particles without interactions. For particles with interactions, the exact representation is very challenging.
One way to improve the representation ability is adding a Jastrow factor before the Slater determinant, and many-body correlations are contained in the Jastrow function.
The backflow correlation improves the representation ability by adding positions of other particles into the single-particle orbital~\cite{backflow on position,backflow_wavefunction_1,backflow_wavefunction_2,NN backflow}.
Backflow corrections on wave-functions have been widely used in quantum chemistry~\cite{DNN schrodinger 1,DNN schrodinger 2}, however for strongly-correlated many-body systems such as Fermi-Hubbard models, the precision is not sufficiently high~\cite{NN backflow,HFSD}.
%Previous works focus on wave-functions, and backflow correction are generated by neural network.

Despite adding variational parameters can increase the state representation ability of the variational wave-function~\cite{DNN schrodinger 1,DNN schrodinger 2,NN backflow}, more variational parameters lead to higher optimization difficulty, which is not beneficial to achieve the ground energy~\cite{CNN_J1J2_5}.
In this work, we propose an efficient way to increase the parameter number of the backflow corrected wave-function by tensor representation. Namely, each dimension of the tensor is an independent degree of freedom in the backflow corrected wave-function.
We show that the tensor representation can achieve competitive or even lower energy results comparing to state-of-the-art fermionic PEPS (fPEPS) results for the spinless $t$-$V$ model~\cite{fPEPS}.
For models with spin, we improve the representation ability by considering backflow corrections on fictitious particles with different spins, which leads to non-zero backflow corrections when the particle and the orbital have different spins.
We numerically demonstrate that our method can
achieve energy precision competitive to state-of-the-art RBM results for molecules under the STO-3G basis~\cite{NN_molecules}, and energy precision competitive with or even better than state-of-the-art NN backflow~\cite{NN backflow} and HFSD~\cite{HFSD} results for the Fermi-Hubbard model.

%The wave-function is a function of single sites and orbitals. In contrast, the Green's function dependes on multiple positions and so it directly represents the spatio-temporal correlations between multiple sites.
%The Green's function is important to understand quantum many-body systems because it associates with the response function~\cite{response}, thus it is useful for calculating conductivity~\cite{conductivity1,conductivity2}. For solving the Hubbard model, a wide application of the Green's function is the dynamical mean-field theory~\cite{DMFT}.

This paper is organized as follows:
Sec.\ref{sec:BW} recalls backflow corrections on wave-functions and introduces our method of backflow corrections when the particle and the orbital have different spins.
Sec.\ref{sec:tensor representation} introduces the tensor representation of the backflow corrected wave-function.
Sec.\ref{sec:numerical} presents numerical results of our methods and comparisons with other state-of-the-art methods.
For example the spinless $t$-$V$ model in Sec.\ref{sec:tV},  molecules in STO-3G basis in Sec.\ref{sec:molecules}, and the Fermi-Hubbard model in Sec.\ref{sec:Hubbard}.
Finally, the paper is concluded in Sec.\ref{sec:conclusions}.

\section{Methods}
%1. Mean field + Definition of fictitious coordinate, Wave function backflow
%
%2. NN backflow + The usage of tensor
%
%3. Green's function and superexchange
%
%4. Green's function backflow + the meaning of backflow

%\textcolor{red}{Mean field and definition}

%In the mean field theory and Hartree-Fock calculation, we approximate the state as the ground state of the quadratic Hamiltonian. The corresponding wave function $|\psi_{\text{MF}}\ket$ is represented by the well-known Slater determinant.
%However the uncorrelated mean-field state, such as BCS state, does not give accurate results under strong interaction strengths and in the presence of frustration.

\subsubsection{Backflow corrections of wave-functions}
\label{sec:BW}
\begin{figure}[t]
\includegraphics[width=0.4\columnwidth]{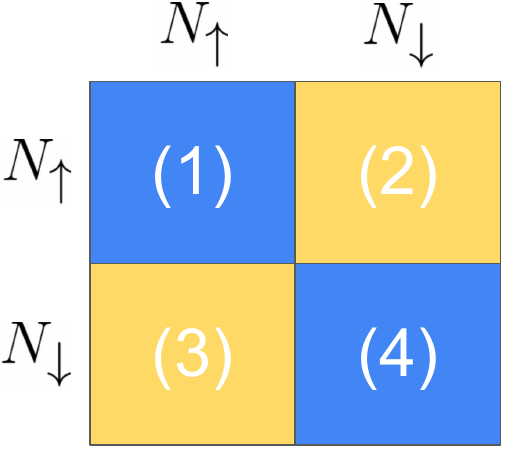}
\caption{The matrix in the Slater determinant for the $N$ particles, with $N_\uparrow$ and $N_\downarrow$ the particle number for spin up and spin down respectively. The spin on the horizontal (vertical) axis denotes the spin of particles (orbitals).}
\label{fig:matrix M}
\end{figure}

The backflow correction is defined on the fictitious coordinate $\mathbf{r}_i^B$ which not only depends on the position $\mathbf{r}_\alpha$ but also depends on positions of other particles~\cite{backflow on position}:
\begin{equation}
	\mathbf{r}_\alpha^B=\mathbf{r}_\alpha+\sum_{\beta}  \eta_{\alpha\beta}[\mathbf{S}] (\mathbf{r}_\beta-\mathbf{r}_\alpha),
	\label{original backflow}
\end{equation}
where $\mathbf{r}_\alpha$ are actual particle positions and $\eta_{\alpha\beta}[\mathbf{S}]$ are variational parameters depending on the many-body state $|\mathbf{S}\rangle$, so to create a return flow of particles.

The backflow corrected single-particle orbital for a spin $\phi^B_{k,\sigma}$ is constructed by a linear combination of eigenstates of the mean-field Hamiltonian $\phi_{k,\sigma}$~\cite{backflow_wavefunction_1,backflow_wavefunction_2,NN backflow}:
\begin{equation}
	\phi^B_{k\sigma}(\mathbf{r}_{i,\sigma})=\phi_{k\sigma}(\mathbf{r}_{i,\sigma})+\sum_{j}c_{ij}[\mathbf{S}]\phi_{k\sigma}(\mathbf{r}_{j,\sigma}),
	\label{BW1}
\end{equation}
where $c_{ij}[\mathbf{S}]$ is a variational coefficient depending on the many-body configuration $|\mathbf{S}\rangle$.
The orbital is $\phi_{k\sigma}(\mathbf{r}_{i,\sigma})=\langle 0|\hat{c}_{i,\sigma}|\phi_{k\sigma}\rangle$, where $\hat{c}_{i,\sigma}$ is the annihilation operator on the $i$-th site with spin value $\sigma$.

Based on Eq.(\ref{BW1}), backflow corrections are performed on positions with an identical spin to the orbital, meanwhile the wave-function is the product of two Slater determinants~\cite{backflow_wavefunction_1,backflow_wavefunction_2}:
\begin{equation}
	w_1(\mathbf{S})=\det[M^{B,\uparrow}]\det[M^{B,\downarrow}],
	\label{w1}
\end{equation}
with the element of the Slater matrix:
\begin{equation}
	M_{ik}^{B,\sigma}= \phi_{k\sigma}^B(r_{i\sigma}).
\end{equation}

%For the Hamiltonian with couplings of spins, particles with different spins are correlated in the ground state.
%In the backflow correction denoted by Eq.(\ref{BW1}), orbitals with different spins are un-correlated, which limits the state representation ability.

To improve the representation ability for the Hamiltonian with couplings of spins, we consider backflow corrections by
the fictitious coordinate for one spin $\mathbf{r}_{i,\sigma_i}^B$ depending on positions of other particles with different spins:
\begin{equation}
	\mathbf{r}_{i,\sigma_i}^B=\mathbf{r}_{i,\sigma_i}+\sum_{j}  \eta_{ij}[\mathbf{S}] \sum_{\sigma_j=\pm 1}(\mathbf{r}_{j,\sigma_j}-\mathbf{r}_{i,\sigma_i}),
	\label{original backflow with spin}
\end{equation}
%therefore particles with different spins are correlated by the summation of $\sigma_j=\pm 1$.
therefore, the backflow correction for one particle at position $\mathbf{r}_i$ with the spin $\sigma_i$ are performed on fictitious particles at positions $\mathbf{r}_j$ with different spins of $\sigma_j$.

Meanwhile, the backflow corrected single-particle orbital for a spin is constructed similarly to Eq.(\ref{BW1}), except the summation on particle spins $\sigma_j$:
\begin{equation}
	\phi^B_{k\sigma_k}(\mathbf{r}_{i,\sigma_i})=\phi_{k\sigma_k}(\mathbf{r}_{i,\sigma_i})+\sum_{j}c_{ij}[\mathbf{S}]\sum_{\sigma_j=\pm 1}\phi_{k\sigma_k}(\mathbf{r}_{j,\sigma_j}),
	\label{BW2}
\end{equation}
where $c_{ij}[\mathbf{S}]$ are variational coefficients depending on the many-body configuration $|\mathbf{S}\rangle$.
The total spin is conserved as $(N_\uparrow-N_\downarrow)/2$, where $N_\uparrow$($N_\downarrow$) is the particle number for spin up(down), as the orbital $\phi_{k\sigma_k}(\mathbf{r}_{i,\sigma_i})$ is zero when $\sigma_k\neq\sigma_i$ and non-zero when $\sigma_k=\sigma_i$.

%Because of the summation on $\sigma_j$ in Eq.(\ref{BW2}), there are non-zero backflow correction terms for arbitrary configurations of the orbital spin $\sigma_k$ and the particle spin $\sigma_i$.
The wave-function is represented by the Slater determinant of a $N\times N$ matrix,
\begin{equation}
        w_2(\mathbf{S})=\det[M^{B}],
        \label{w2}
\end{equation}
where $M_{ik}^B=\phi^B_{k\sigma_k}(\mathbf{r}_{i,\sigma_i})$ and $N=N_\uparrow+N_\downarrow$ is the total particle number.

In this paper, backflow corrections of Eq.(\ref{BW1}) and Eq.(\ref{BW2}) are denoted as BW1 and BW2, respectively.
In the case of BW1, it requires the particle spin equals to the orbital spin in order to achieve non-zero backflow correction terms.
However in the case of BW2, because of the summation on $\sigma_j$, there are non-zero backflow correction terms for arbitrary configurations of the orbital spin $\sigma_k$ and the particle spin $\sigma_i$.

Therefore, the major difference between BW1 and BW2 is in the matrix in the Slater determinant, and the matrix is depicted in Fig.(\ref{fig:matrix M}).
In the matrix, the particle and the orbital have the identical spin in submatrices (1)(4), meanwhile in submatrices (2)(3) the particle and the orbital have opposite spins.
For BW1,  submatrices (2)(3) are undefined, meanwhile no backflow corrections are performed.
For BW2, there are non-zero backflow corrections in all submatrices.

%Backflow correction terms can be generated by a NN with the input of the NN the many-body configuration~\cite{}.

%and it can be represented by the expectation of
%$\hat{c}_{1'}^\dagger\cdots\hat{c}_{N'}^\dagger$ for the configuration $|\mathbf{S}'\rangle$ multiplied by
%$\hat{c}_1\cdots\hat{c}_N$ for the configuration $|\mathbf{S}\rangle$.
%The operator $\hat{c}_{i'}^\dagger$ creates the $i$-th particle in the configuration $|\mathbf{S}'\rangle$. Namely,  See Appendix. A for more explicit details.

%\textcolor{red}{Tensor and NN backflow}

%In recent work, Neural Network has been applied to dress the single particle orbitals in the Slater determinant. They replace the elements in the determinant with configuration dependent neural network, which represent a backflow transformation and significantly decreases the relative error. The single particle orbitals with backflow is given by
%\begin{equation}
%	\phi^B_{k,\sigma}(\mathbf{r}_i)=\phi_{k,\sigma}(\mathbf{r}_i)+a_{ki,\sigma}^{\text{NN}}(\mathbf{r},\mathbf{S})
%\end{equation}
%where $a_{k,i}^{\text{NN}}(\mathbf{r})$ is the output neuron of the Neural network.

\subsubsection{Tensor representation of backflow corrections}
\label{sec:tensor representation}
We evaluate state representation abilities of BW1 and BW2 by using tensor representations of backflow corrected wave-functions.
Each dimension of the tensor is an independent degree of freedom in the backflow corrected wave-function denoted by Eq.(\ref{BW1}) and Eq.(\ref{BW2}).

The tensor has representation ability beyond what is provided by the linear combination of eigenstates of the mean-field Hamiltonian, as it includes much more variational parameters.
Comparing to the backflow corrections for specific configurations of $\mathbf{s}(\mathbf{r}_i)$ and $\mathbf{s}(\mathbf{r}_j)$~\cite{backflow_wavefunction_1,backflow_wavefunction_2}, it includes all possible configurations.
Furthermore, it can represent higher order correlations, such as two-body correlations in the calculation of self-energy based on the diagrammatic perturbation expansion~\cite{backflow_wavefunction_2,perturbation}.

Here we consider coefficients $c_{ij}$ in Eq.(\ref{BW1}) and Eq.(\ref{BW2}) depend on local configurations $\mathbf{s}(\mathbf{r}_i)$ and $\mathbf{s}(\mathbf{r}_j)$ instead of $|\mathbf{S}\rangle$ for simplicity~\cite{backflow_wavefunction_1,backflow_wavefunction_2}.
Independent degrees of freedom in both BW1 and BW2 are the position $\mathbf{r}_{i,\sigma}$, the orbital number of $\phi_{k\sigma}$, the summation index $j$, configurations $\mathbf{s}(\mathbf{r}_i)$ and $\mathbf{s}(\mathbf{r}_j)$.
Therefore the total dimension of the tensor is:
\begin{equation}
	[M,N,d,Q,d],
	\label{tensor dimension}
\end{equation}
where the dimension $M$ equals to the site number, and the dimension $N$ equals to the total particle number.
%Because the configuration $|\mathbf{S}'\rangle$ is fixed, in $|\mathbf{S}'\rangle$ the position and the spin of the particle are treated as an index of the particle, with the dimension is the particle number, namely $j'\in[1,\cdots,N]$.
The first and the second $d$ are for configurations $\mathbf{s}(\mathbf{r}_i)$ and $\mathbf{s}(\mathbf{r}_k)$, respectively.
$Q$ denotes the index of $i$ as well as indexes in the summation of $j$ considered in either Eq.(\ref{BW1}) and Eq.(\ref{BW2}).

The matrix element in the Slater determinant is assigned by indexing the tensor:
\begin{equation}
	M_{ik}^B=g[i,k,\mathbf{s}(\mathbf{r}_i),i,\mathbf{s}(\mathbf{r}_k)]
	+\sum_{\langle q,i \rangle}g[i,k,\mathbf{s}(\mathbf{r}_i),q,\mathbf{s}(\mathbf{r}_q)],
	\label{Mij}
\end{equation}
where $g$ is the tensor representation of the wave-function, with the dimension defined in Eq.(\ref{tensor dimension}). The forward calculation generates the wave-function coefficient $w(\mathbf{S})$ defined by Eq.(\ref{w1}) for BW1 and Eq.(\ref{w2}) for BW2.

In the variational Monte-Carlo (VMC), forward and backward calculations of the wave-function coefficient $w(\mathbf{S})$ are necessary.
The forward is achieved by calculating the Slater determinant of the $N\times N$ matrix, and
the backward is achieved by inverting the matrix. Details for backward calculations are in the Appendix~\ref{backward}.

%In this work, we introduce the tensor as the backflow term to correct the wave function. Using tensor instead of directly training the variational coefficient $c_{i,j}$ can greatly decrease the difficulty of training. It will avoid some local minima and represent all possible virtual hopping and super-exchange. In the meantime, compressing these variational parameter into one tensor will increase the represent-ability of the ansatz .

%And we add the backflow correlation into the one-body Green's function. The Green's function naturally represents the spatio-temporal correlations between multiple sites. From the Hamiltonian, the energy expectation of a hopping term is a one-body correlation, and that of an interaction term is a two-body correlation\cite{energy_1}.

%\textcolor{red}{Definition and density matrix}

\section{Numerical Investigations}
\label{sec:numerical}
In this section we numerically demonstrate that backflow corrections under the tensor representation have strong representation abilities.
We benchmark on three types of models:
(1), the spinless fermionic $t$-$V$ model on the square lattice with OBC.
(2), several molecules under the STO-3G basis.
(3), the spinful Fermi-Hubbard model on rectangular lattices with PBC and cylindrical boundary condition (CBC).

For the spinless $t$-$V$ model, backflow corrections can achieve state-of-the-art energy results comparing to the PEPS.
For molecules under STO-3G basis, BW2 has better precision than BW1, and energies obtained by BW2 are competitive to state-of-the-art results.
For the Fermi-Hubbard model, both BW1 and BW2 achieve competitive or even lower energy results comparing to state-of-the-art methods like NN backflow or HFSD. Furthermore, energies obtained by BW2 are lower comparing to BW1, on finite sized lattices.

\subsubsection{Optimization methods}
\label{sec:optimization methods}
The wave-function is first optimized by the VMC, then further optimized by a Lanczos step.
In VMC, the energy and the $\alpha$-th parameter's gradient are evaluated through the Markov-Chain-Monte-Carlo (MCMC) process~\cite{fPEPS,PEPS_liu}:
\begin{equation}
	\begin{split}
	&E=\langle E_{\mathrm{loc}}\rangle\\
	&G^\alpha= 2\langle E_{\mathrm{loc}}O_{\mathrm{loc}}^\alpha \rangle
	-2\langle E_{\mathrm{loc}} \rangle\langle O_{\mathrm{loc}}^\alpha \rangle,
		\label{E and G}
	\end{split}
\end{equation}
where the local energy is $E_\mathrm{loc}(\mathbf{S})=\sum_{\mathbf{S}'}\frac{w(\mathbf{S}')}{w(\mathbf{S})}\langle \mathbf{S}'| \hat{H}|\mathbf{S} \rangle$, the $O_{\mathrm{loc}}(\mathbf{S})^\alpha=\frac{1}{w(\mathbf{S})}\frac{\partial w(\mathbf{S})}{\partial\alpha}$, and $\langle \cdots \rangle$ denotes the average on MCMC samples.

The variational parameters are updated according to the gradient descent method.
Here we only adopt the first-order gradient descent due to the low optimization difficulty of the tensor representation. Because of the limited MC sample number, we take the sign of the gradient and apply a constant step size $\delta$: $\alpha'=\alpha-\delta\mathrm{sgn}(G^\alpha)$. Such parameter updating scheme has been successfully used in optimizing high dimensional tensors like the PEPS~\cite{fPEPS,PEPS_GO_1,PEPS_GO_2}.

A Lanczos step further improves the representation ability of wave-function $|\Psi_{p=0}\rangle$ by considering an additional wave-function $|\Psi_{p=0}^\perp\rangle$ orthogonal to $|\Psi_{p=0}\rangle$~\cite{CNN_J1J2_5,Lanczos}:
\begin{equation}
	|\Psi_{p=1}\rangle=A|\Psi_{p=0}\rangle+B|\Psi_{p=0}^{\perp}\rangle,
\end{equation}
where $A$ and $B$ are parameters to be determined, and $|\Psi_{p=0}\rangle$ is the wave-function obtained after the VMC. The orthogonal wave-function is built by $|\Psi_{p}^\perp\rangle=\frac{1}{\sigma_p}(\hat{H}-E_p)|\Psi_{p}\rangle$, where the energy expectation $E_{p}=\langle \Psi_{p}|\hat{H}|\Psi_{p} \rangle$ and the variance $\sigma_{p}^2=\langle\Psi_{p}| (\hat{H}-E_p)^2|\Psi_{p} \rangle$.

\subsubsection{Spinless t-V model}
\label{sec:tV}
\begin{table}[]
	\caption{Comparisons of energies(per site) for the spinless $t-V$ model on $10\times 10$ square lattice under OBC, total particle number is 50. $p$=0($p$=1) denotes the wave-function before(after) one Lanczos step. Reference energies are obtained by the fPEPS method~\cite{fPEPS}.}
\begin{tabular}{|c|ccc|c|}
\hline
$V$  & HF      & \ \ \ \ $p$=0\ \ \ \   & \ \ \ \ $p$=1\ \ \ \    & \ fPEPS\  \\ \hline
0.45 & -0.6103 & -0.6132 & -0.6134 & -0.6129   \\ \cline{1-1}
1    & -0.4561 & -0.4617 & -0.4620 & -0.4620   \\ \cline{1-1}
2    & -0.2961 & -0.2997 & -0.2999 & -0.2999   \\ \hline
\end{tabular}
	\label{tVenergies}
\end{table}
\begin{figure}[]
\includegraphics[width=0.9\columnwidth]{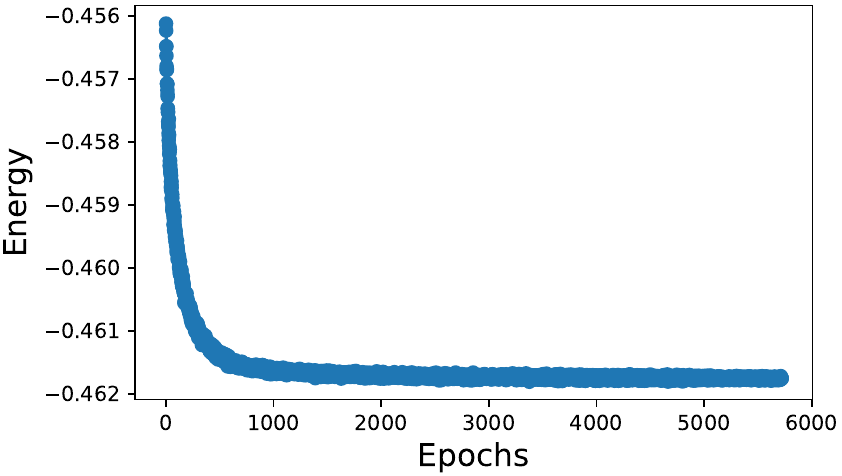}
\caption{The energy convergence of the first-order gradient descent for the spinless $t-V$ model with $V/t=1$ on the $10\times10$ lattice under OBC.
The MC sample number 128000, and the parameter updating step size $\delta=5\times 10^{-4}$.
The converged energy per site is -0.4617, while the reference energy obtained by fPEPS is -0.4620.}
\label{fig:converge}
\end{figure}

The Hamiltonian of the spinless $t$-$V$ model reads:
\begin{equation}
	\hat{H}=-t\sum_{\langle i,j\rangle}(\hat{c}_i^{\dagger} \hat{c}_j+h.c.)+V\sum_{\langle i,j\rangle}\hat{n}_{i}\hat{n}_{j},
\end{equation}
where $t$ is the hopping strength and $V$ is the interaction strength between nearest neighbours. $\hat{c}_i^\dagger$ ($\hat{c}_i$) creates (destroys) a particle on the $i$-th site, and the particle number operator $\hat{n}_i=\hat{c}_i^\dagger\hat{c}_i$. We set $t$=1 through our investigations.
The maximal occupation per site is $n_i=1$ for the spinless $t-V$ model. We investigate the half-filling case $n=1/2$ so that the particle number is half of the total site number, on the $10\times10$ square lattice.

We benchmark on lattices with OBC to compare with the state-of-the-art fPEPS method~\cite{fPEPS}. The fPEPS is optimized by the imaginary-time-evolution method called the simple-update and then by the gradient descending method~\cite{fPEPS,PEPS_liu,PEPS_GO_1,PEPS_GO_2}, thus the fPEPS gives high precision energy references.
For the fPEPS, the bond dimension is 8 and the truncation dimension is 32.

For the spinless model, backflow corrections are performed for one orbital based on Eq.(\ref{original backflow}).
Energy comparisons under different interaction strengths $V$ are denoted in Table.(\ref{tVenergies}).
In the table, results of HF are achieved by representing the spinless HF orbital $\phi_k(\mathbf{r}_i)$ by a tensor with the dimension of $[M, N]$, where $M$ is the site number and $N$ is the total particle number.
$p$=0 is the result obtained by the tensor representation of backflow corrections, and $p$=1 is the result of one Lanczos step for the $p$=0 wave-function.
Backflow corrections are considered on nearest neighbours of $\mathbf{r}_i$, namely the dimension $Q$ defined in Eq.(\ref{tensor dimension}) includes the site $\mathbf{r}_{i}$ as well as its nearest neighbours, therefore $Q=5$.

Comparing to $p$=1 results, relative errors of HF are in the magnitude of $10^{-3}$ for $V=0.45$ and the magnitude of $10^{-2}$ for $V=1,2$, meanwhile backflow corrections decrease relative errors to the magnitude of $10^{-4}$ for all cases.
From the table, both $p$=0 and $p$=1 have energy precision competitive to the fPEPS.

%Near the phase transition region $V$=0.45, both $p$=0 and $p$=1 have better energy precision than the PEPS. For $V$=1 and $V$=2, both $p$=0 and $p$=1 have energy precision competitive to the fPEPS.

%The convergence is also revealed by the energy variance, as the energy variance of the ground state is in principle zero~\cite{Lanczos}. For $p$=0 wave-functions, the variance per site definded as $\sigma_p^2/100$ is in the magnitude of $10^{-3}$ for $V=0.45,1$ and the magnitude of $10^{-4}$ for $V=2$.

The energy convergence of VMC for the spinless $t-V$ model with $V/t=1$ is depicted by Fig.\ref{fig:converge}.
The initial energy -0.4561 is from the HF state. After HF is converged, we continue the optimization by adding backflow corrections, within the initial energy from the HF.
The energy converges smoothly after backflow corrections are added.
The parameter updating step size is $\delta=5\times 10^{-4}$ and the MC sample number for each step is 128000.
The interval between two MC samples is the total site number. The converged energy per site is -0.4617, meanwhile the reference energy obtained by the fPEPS is -0.4620.

\subsubsection{Molecules on STO-3G basis}
\label{sec:molecules}
\begin{table}[]
	\caption{Comparisons of energies achieved by several methods for molecules under the STO-3G basis. $M$ is the equivalent site number and $N$ is the particle number. All energy results achieved by BW1 and BW2 are evaluated by $p$=0 wave-functions. Energy results from CCSD(T) and RBM are from the literature~\cite{NN_molecules}.}
\begin{tabular}{|c|cc|cc|c|c|}
\hline
Molecule &\  $M$& $N$\ &\ \ BW1\ \         &\ \ BW2\ \         &\ CCSD(T)\    &\ \ RBM\ \     \\ \hline
H2O      & 14& 10&-75.0201  & -75.0221  & -75.0231  & -75.0232  \\ \cline{1-1}
NH3      & 16& 10&-55.5181  & -55.5274  & -55.5281  & -55.5277  \\ \cline{1-1}
C2       & 20& 12&-74.6808  & -74.6865  & -74.6876  & -74.6892  \\ \cline{1-1}
N2       & 20& 14&-107.6585 & -107.6742 & -107.6738 & -107.6767 \\ \hline
\end{tabular}
	\label{molecules}
\end{table}

The Hamiltonian for molecules in the second quantization form is:
\begin{equation}
	\hat{H}=\sum_{ij}t_{ij}\hat{c}_i^{\dagger}\hat{c}_j+\sum_{ijkm}u_{ijkm}\hat{c}_i^\dagger\hat{c}_k^\dagger\hat{c}_j\hat{c}_m,
	\label{H_molecule}
\end{equation}
here label $i$ denotes the fermionic mode, with $t_{ij}$ the one-body interaction and $u_{ijkm}$ the two-body interaction. $\hat{c}_{i}^\dagger$ ($\hat{c}_{i}$) creates (destroys) a particle on the $i$-th fermionic mode. %In the Hamiltonian, the effective site number is twice as the total orbital number because of the spin.
The structure of a molecule under STO-3G is obtained from the literature~\cite{NN_molecules}, and we use the software package \texttt{PySCF}~\cite{pyscf} to generate coefficents of $t_{ij}$ and $u_{ijkm}$, with the maximal iteration number of 500.
We first optimize a HF state for each molecule without two-body interactions,
then continue the optimization by adding backflow corrections from the HF.

A HF orbital with spin $\phi_{k\sigma}(\mathbf{r}_{i,\sigma})$ is represented by a tensor with the dimension of $[M,N,d]$,
where $M$ is the equivalent site number and $N$ is the total particle number.
The equivalent site number is twice as the total orbital number because of the spin on-site.
Configurations on each equivalent site are occupation and non-occupation, thus $d=2$.
For backflow corrections, we consider backflow terms from all equivalent sites, thus the dimension of the tensor is defined by Eq.(\ref{tensor dimension}) with $Q=M$.

A challenge for solving molecules is the locality of the ground state in the total Hilbert space, as the ground state is nearly classical.
Therefore, the optimization is easily to get stuck on local minima especially when using a large MC sample number in initial optimization steps.

To avoid local optimizations, we stop the optimization after hundreds of optimziation steps, then continue the optimization by enforcing the MCMC starting from the configuration based on the Pauli exclusion principle.
For each case, we start from 44800 MC samples in initial optimization steps. After one thousand optimizations, we use roughly 70000 MC samples for each optimization step, with the interval between two MC samples equal to the number of equivalent sites.

Energy comparisons are denoted by Tab.(\ref{molecules}).
Each energy of both BW1 and BW2 is obtained by the average of the last 20 optimization steps.
From the table, energies precision achieved by BW1 is lower than that achieved by BW2, and energy results obtained by BW2 are competitive to state-of-the-art results obtained by CCSD(T) and RBM~\cite{NN_molecules}.
From the table, RBM achieve better energy precision than BW.
RBM is efficient for molecules under the STO-3G basis because the wave-function is easy to be represented and do not demand the full expressibility of BW.

\subsubsection{Fermi-Hubbard model}
\label{sec:Hubbard}
\begin{figure}[t]
\includegraphics[width=0.9\columnwidth]{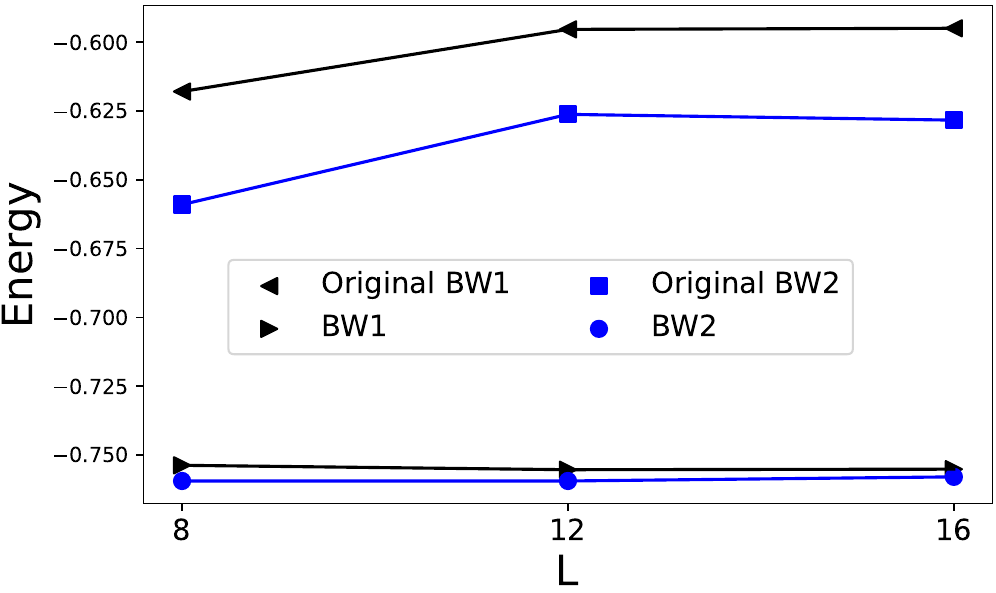}
\caption{Comparisons of energies(per site) between the original backflow and the single tensor representation for the Fermi-Hubbard model of $n=0.875$, on the $4\times L$ lattice with PBC. Each energy is evaluated by the $p=0$ wave-function. }
\label{fig:tensor_comparison}
\end{figure}
The Hamiltonian of the Fermi-Hubbard model is:
\begin{equation}
	\hat{H} = -t \sum_{\langle ij\rangle,\sigma}(\hat{c}_{i\sigma}^\dagger \hat{c}_{j\sigma}+h.c.)+U \sum_i \hat{n}_{i\uparrow} \hat{n}_{j\downarrow},
	\label{H}
\end{equation}
where $t$ is the hopping strength and $U$ is the strength of on-site interactions. $\hat{c}_{i\sigma}^\dagger(\hat{c}_{i\sigma})$ creates(destroys) a particle of spin $\sigma$ on $i$-th site, and the particle number operator $\hat{n}_{i\sigma}=\hat{c}_{i\sigma}^\dagger \hat{c}_{i\sigma}$. For the Hubbard model with spin, double occupations are allowed. We set $t$=1 through our investigations.
In each optimization step, there are roughly 44800 MC samples for calculating gradients, and the interval between two samples is the lattice size.

We first demonstrate the advantage of our single tensor representation by comparing to the original backflow.
In this work, we represent the original form of the wave-function defined in either BW1 or BW2 by two separate tensors.
The first tensor represents coefficients $c_{ij}$ with the dimension of $[M,Q,d,d]$, where $M$ is the site number, $Q=5$ denotes the position $\mathbf{r}_i$ as well as its nearest neighbours, $d=4$ denotes degrees of freedom per site.
The second tensor represents the HF orbital $\phi_k(\mathbf{r}_{i,\sigma})$ with the dimension of $[M,N,2]$, where $N$ is the total particle number and the dimension of 2 denotes the spin.

For representing the backflow corrected wave-function by a single tensor,
the dimension of the tensor is $[M,N,2,d/2,Q,d]$, where $d=4$ denotes degrees of freedom per site. In the tensor, the dimension of 2 is for $\sigma_i$, and the dimension of $d/2$ denotes whether there is double occupation on $\mathbf{r}_i$.
Comparing to the dimension defined in Eq.(\ref{tensor dimension}), we divide the first $d$ in order to distinguish the double occupation.
$Q=5$ denotes the position $\mathbf{r}_i$ as well as its nearest neighbours.

Fig.(\ref{fig:tensor_comparison}) denotes energy comparisons between the original backflow and the single tensor representation on the $4\times L$ lattice with PBC, the filling of the Hubbard model is $n=0.875$.
Each energy result is evaluated by the $p=0$ wave-function.
From the figure, for either BW1 or BW2, the original backflow has much worse energy precision than the single tensor representation.
For either the original backflow or the single tensor representation, BW2 has better energy precision than BW1.
The HF wave-function is represented by the tensor with the dimension of $[M,N,2]$, and it achieves -0.5330 for $L=8$, -0.5398 for $L=12$ and -0.5658 for $L=16$. Thus backflow corrections achieve better energy precision than HF.

%From our investigations, the initial state is not vital for the energy convergence of BG, because differences of converged energies with HF initial states under different values of $U$ are negligible. The energy will converge even without a HF initial state by a fully random initialization. However using HF initial states will reduce the optimization time.

\begin{table}[]
	\caption{Comparisons of energies(per site) for the Fermi-Hubbard model on rectangular lattices with PBC. Reference energies for $n=1$ are from the AFQMC~\cite{AFQMC_2}. For $n=0.875$, the Ref.1 are from NN backflow~\cite{NN backflow} and the Ref.2 are from HFSD~\cite{HFSD}. All energy results of BW1 and BW2 are evaluated by $p$=1 wave-functions. }
\begin{tabular}{|c|c|cc|c|c|}
\hline
$n$ &
  Lattice Size &
  \ \ \  BW1\ \ {} &
  \ \  BW2\ \ \ {} &
  \ \ \  Ref.1\ \ \ {} &
  \ \ \  Ref.2\ \ \ {} \\ \hline
\multirow{3}{*}{1} & $6\times 6$   & -0.5186 & -0.5257 & -0.5278 & --      \\ \cline{2-2}
                   & $8\times 8$   & -0.5188 & -0.5241 & -0.5263 & --      \\ \cline{2-2}
                   & $10\times 10$ & -0.5181 & -0.5230 & -0.5254 & --      \\ \hline
\multirow{5}{*}{0.875} & $4\times 8$   & -0.7591 & -0.7633 & -0.755  & -0.7633 \\ \cline{2-2}
                   & $4\times 12$  & -0.7608 & -0.7636 & -0.746  & --      \\ \cline{2-2}
                   & $4\times 16$  & -0.7597 & -0.7618 & -0.746  & -0.753  \\ \cline{2-2}
                   & $4\times 20$  & -0.7566 & -0.7591 & --      & --      \\ \cline{2-2}
                   & $4\times 24$  & -0.7577 & -0.7595 & --      & --      \\ \hline
\end{tabular}
	\label{Hubbard energies}
\end{table}
\begin{figure}[t]
\includegraphics[width=0.95\columnwidth]{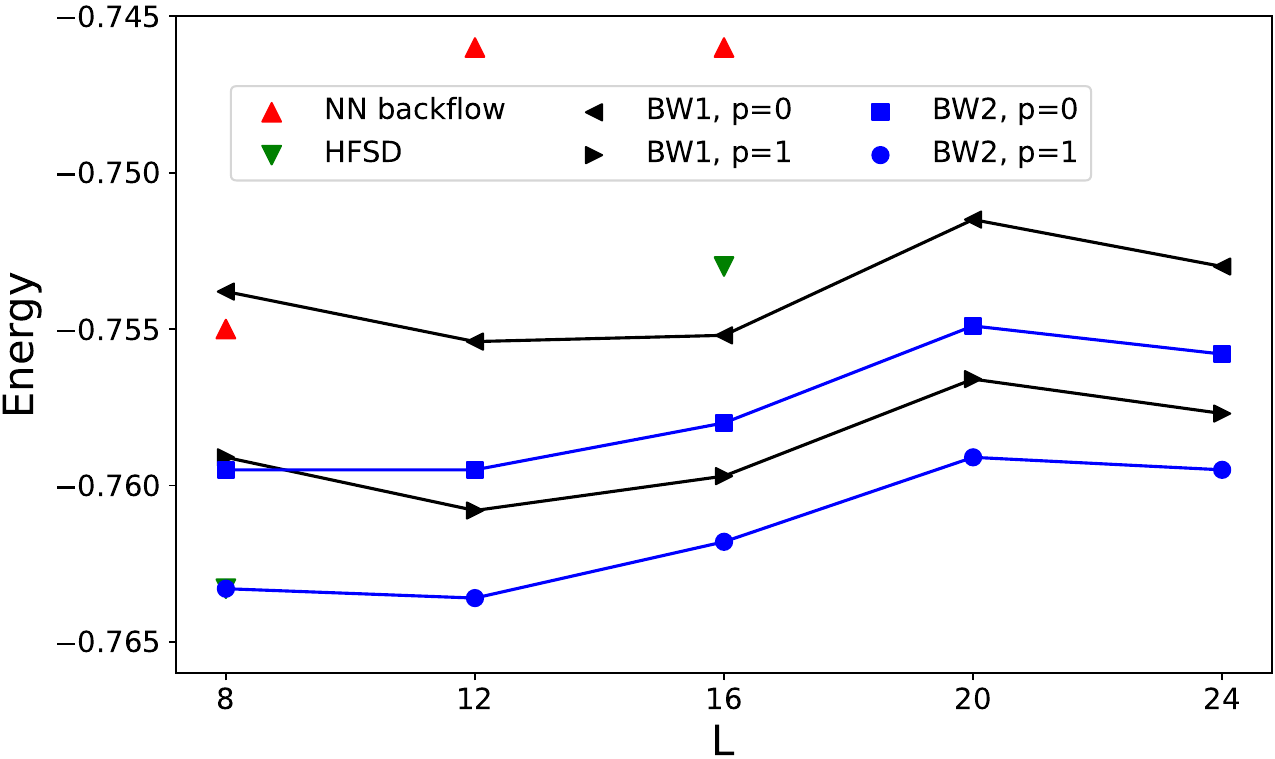}
\caption{Energy comparisons for Fermi-Hubbard model on $4\times L$ lattices under PBC. The filling is $n=0.875$. Red upper triangles denote NN backflow, Green's lower triangles denote HDFS. BW1 energy results of $p$=0($p=1$) are denoted by left(right) triangles. BW2 energy results of $p$=0($p$=1) are denoted by squares(circles). }
\label{fig:energy_comparison}
\end{figure}
Tab.(\ref{Hubbard energies}) denotes energy comparisons of BW1, BW2 and other state-of-the-art methods on square lattices under the PBC.
For results obtained by BW1 and BW2 in the table, we first optimize a HF state under $U$=0, and then adding backflow corrections and continuing optimizations,
each energy value is evaluated by the $p=1$ wave-function using 48000 MC samples, and the interval between two MC samples is the lattice size.

%Energy results of BW is achieved by omitting area (2)(3) in the matrix denoted by Fig.(\ref{fig:matrix M}).
For cases of half filling $n=1$, the reference energies are from AFQMC~\cite{AFQMC_2}.
On the $6\times 6$ lattice, the relative error is $1.7\times 10^{-2}$ for BW1 and $4.0\times 10^{-3}$ for BW2.
On the $8\times 8$ lattice, the relative error is $1.4\times 10^{-2}$ for BW1 and $4.2\times 10^{-3}$ for BW2.
On the $10\times 10$ lattice, the relative error is $1.4\times 10^{-2}$ for BW1 and $4.5\times 10^{-3}$ for BW2.
From the table, BW2 has better energy precisioin than BW1.

For the more challenging case of 1/8 doping $n=0.875$,
Ref.1 and Ref.2 in Tab.(\ref{Hubbard energies}) are from NN backflow~\cite{NN backflow} and HFSD~\cite{HFSD}, respectively.
To clearly compare energies, energies for the filling $n=0.875$ on $4\times L$ lattices with PBC are depicted in Fig.(\ref{fig:energy_comparison}).
For $p$=0 results, BW1 achieves satisfactory energies comparing to NN backflow, however the energy precision is not competitive to HFSD on the $4\times 8$ lattice.
However on the $4\times 16$ lattice, BW1 achieves lower energy than HFSD.
For all lattice sizes, BW2 achieve better precision than BW1, and BW2 achieve lower energies than those from NN backflow.
Lanczos can significantly improve the energy precision for all cases.
For $p$=1 cases, both BW1 and BW2 achieve better energy precision than NN backflow, and BW2 achieve better energy precision than HFSD.

Comparing to the DMRG energy -0.7659 for system size $4\times\infty$(open, PBC)~\cite{NN backflow}, our energy -0.7595 or $4\times 24$ is still higher.
Because only nearest neighbours of $\mathbf{r}_i$ are considered in our backflow corrections, our energy precision can be improved by considering further backflow corrections.

\begin{figure}[t]
\includegraphics[width=1\columnwidth]{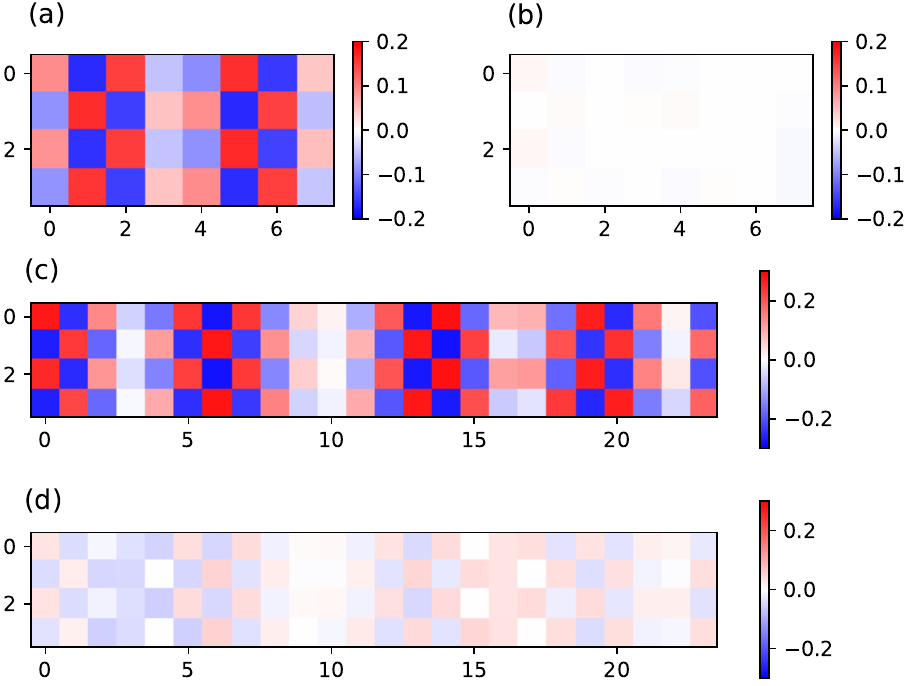}
\caption{The spin density of the Fermi-Hubbard model of filling $n=0.875$ and $U$=8 with PBC, evaluated on the $p$=1 wave-function.
On the $4\times 8$ lattice, the spin density obtained by BW1 and BW2 are denoted by (a) and (b), respectively.
The spin density on the $4\times 24$ lattice obtained by BW1 and BW2 are denoted by (c) and (d), respectively.}
\label{fig:SD and CD}
\end{figure}
Besides comparisons on energies, detailed differences between BW1 and BW2 are depicted in Fig.(\ref{fig:SD and CD}).
In the figure, the spin density is defined as the average spin value each site: $\langle \mathbf{S}_i\rangle$.
%and the charge density is defined as the average particle number each site: $\langle n_i^\uparrow+n_i^\downarrow\rangle$.
Because of the PBC, the spin density is ideally uniform, and the spin density is ideally zero due to the total spin of the ground state is zero.

On the $4\times 8$ lattice, from Fig.(\ref{fig:SD and CD})(a)(b), the ground state obtained by BW2 has a more uniform spin density than that obtained by BW1.
%From Fig.(\ref{fig:SD and CD})(c)(d), the charge density obtained by either BW1 or BW2 is nearly uniform.
On the $4\times 24$ lattice, Fig.(\ref{fig:SD and CD})(c)(d) depict the spin density for BW1 and BW2, respectively.
From the figure, the ground state achieved by BW2 has a more uniform spin density than that achieved by BW1.
It is notable that the ground energy achieved by BW2 is only $2.4\times 10^{-3}$ lower than that achieved by BW1, thus the BW2 has more representation ability than BW1.
%Both spin densities achieved by BW1 and BW2 have the stripe order with the period of $\lambda=8$, which match results obtained by the HFSD~\cite{HFSD}, and BW2 has a more uniform spin density than BW1.

%Fig.(\ref{fig:stripe})(a) depicts the spin density $\langle \sigma_\mathbf{i} \rangle/2$ on the $4\times 16$ lattice with PBC, evaluated on the $p$=1 wave-function. The stripe order with the period $\lambda=8$ is achieved, which is consistent with previous studies~\cite{Hubbard_2}. Comparing to the results achieved by HFSD~\cite{HFSD}, our stripe pattern has shifted one site along the longer boundary because of the PBC.

\begin{figure*}[t]
\includegraphics[width=1.4\columnwidth]{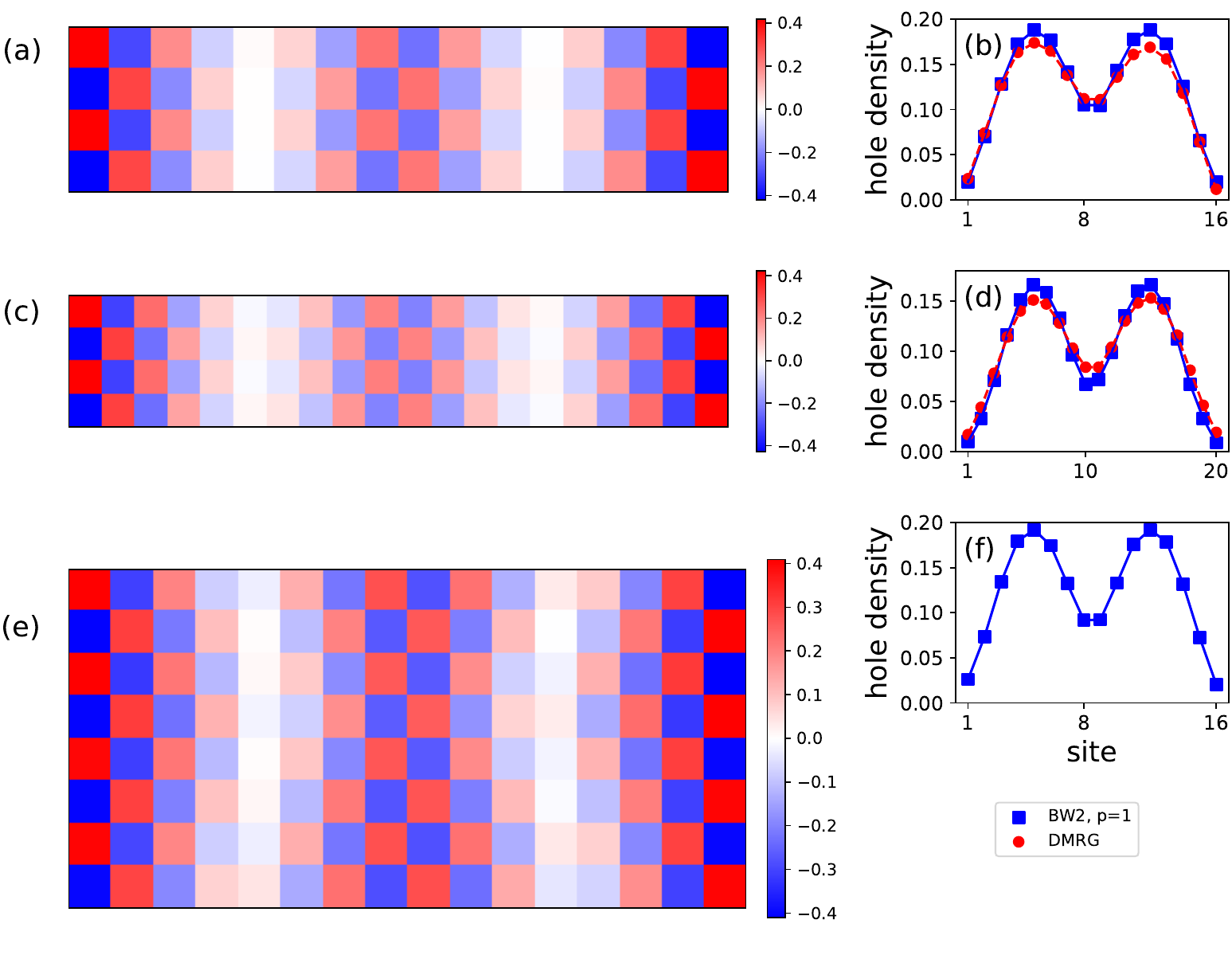}
\caption{The spin density and the hole density achieved by BW2 on rectangular lattices under CBC(a)(b)(c)(d) and PBC(e)(f), with a pinning field applied on both short edges. (a)(b) depict the $4\times 16$ lattice, with the filling $n=0.875$. (c)(d) depict the $4\times 20$ lattice, with the filling $n=0.9$. (e)(f) depict the $8\times 16$ lattice, with the filling $n=0.875$. }
\label{fig:stripe}
\end{figure*}
Based on the results on PBC, we investigate BW2 on cases that ground states are supposed to have stripe orders, such as rectangular lattices under CBC in previous literatures~\cite{AFQMC_1,AFQMC_3}. The boundary conditions are open along the shorter boundary($x$) and periodic along the longer boundary($y$). To break degeneracy from translational symmetry, a pinning field is applied on both shorter boundaries: $v_{i\uparrow}=-v_{i\downarrow}=(-1)^{i_x+i_y}v_0$ for $i_y=1$ and $i_y=L_y$.

Fig.(\ref{fig:stripe})(a)(b)(c)(d) depict the spin density and the hole density on CBC, evaluated on the $p$=1 wave-function.
Fig.(\ref{fig:stripe})(a)(b) depict the $4\times 16$ lattice and filling $n=0.875$, $U=8$ with the pinning field strength $v_0=0.25$. The reference energy by DMRG is -0.7713, and the BW2 achieves energy -0.7640 for $p$=0 and -0.7678 for $p$=1. The relative energy error is $9.4\times 10^{-3}$ for $p$=0 and $4.5\times 10^{-3}$ for $p$=1.
Fig.(\ref{fig:stripe})(c)(d) denote the lattice size $4\times 20$ and filling $n=0.9$, $U=6$ with the pinning field strength $v_0=0.5$.
BW2 achieves the ground energy of -0.8485 for $p$=0 and -0.8516 for $p$=1. Comparing to the energy reported by DMRG -0.8352, BW2 is $1.6\times 10^{-2}$ lower for $p$=0 and $1.9\times 10^{-2}$ lower for $p$=1.
The stripe patterns depicted in Fig.(\ref{fig:stripe})(a)(c) match those from both AFQMC and DMRG~\cite{AFQMC_1,AFQMC_3}.

Furthermore, we benchmark BW2 on lattice as large as $8\times 16$, with PBC on both directions, with filling $n$=0.875 and $U$=8. To break degeneracy from translational symmetry, a pinning field is applied on both shorter boundaries with field strength $v_0=0.25$.
With the interval between two MC samples the lattice size, it takes roughly 2 minutes for one optimization step with a total of 44800 MC samples on 128 AMD EPYC 7742 CPU cores.
The converged energies are -0.7748 for $p=0$ and -0.7784 for $p=1$.
Fig.(\ref{fig:stripe})(e)(f) denote the spin density evaluated on the $p=1$ wave-function and the hole density, respectively.
The spin density pattern matches that on the $4\times 16$ lattice, which demonstrates the valid state representation ability of BW2 on large lattices.

\section{Conclusions}
\label{sec:conclusions}
We show that tensor representations of backflow corrections after a Lanczos optimization have sufficient representation abilities for achieving state-of-the-art ground energies.
Because the tensor representation is easy to optimize, first-order gradient descent is feasible.

For systems with spins, the representation ability can be further improved by considering backflow corrections on different spins, and natually introduce non-zero backflow corrections where the particle and the orbital have opposite spins.
By numerical demonstrations on molecules on STO-3G basis and the finite sized Fermi-Hubbard model, we show that BW2 has a better representation ability than BW1.
Furthermore, we show that BW2 is capable to solve systems on large lattice sizes.
%Our tensor representation has reduntant dimensions, thus the parameter number can be further reduced, which is helpful for calculating larger systems.

For the Fermi-Hubbard model, comparing to the NN backflow~\cite{NN backflow}, the input of the NN is the total many-body configuration $|\mathbf{S}\rangle$.
In either BW1 or BW2, backflow terms are limited as nearest neighbours of the position $\mathbf{r}_i$, and the energy precision can be improved by considering backflow terms with further distances.
Comparing to HFSD~\cite{HFSD}, the HFSD considers a matrix larger than $N\times N$ in the Slater determinant by introducing additional hidden particles. Here in either BW1 or BW2, the size of the matrix is kept as $N\times N$, and the enhanced representation ability is achieved by additional degrees of freedom introduced by backflow correction terms in the $N\times N$ matrix.

In our tensor representations, backflow corrections are performed between two sites, however the representation ability can be in principle improved by considering higher order correlations, thus considering backflow corrections in the perspective of Green's function is feasible.
The representation ability can be improved by increasing variational parameters, thus the application of neural networks based on the tensor representation is feasible.
We hope our work will give some insights on developping numerical methods for solving quantum many-body systems.

\section{Acknowledgement}
X. Liang thanks Giuseppe Carleo, Di Luo, Shiwei Zhang, Ao Chen, Lixin He, Michael Widom and Yang Wang for usefull discussions.
Y.-T. Zhou and Z.-W. Zhou are supported by National Natural Science Foundation of China (Grants No.11974334) and Innovation Program for Quantum Science and Technology (Grant No.2021ZD0301900).
X. Liang is supported by the NSF award number OAC-2139536.
This work used the Bridges-2 system, which is supported by NSF award number OAC-1928147 at the Pittsburgh Supercomputing Center (PSC).
This research was also supported by Texas Advanced Computing Center (TACC) at The University of Texas at Austin and the advanced computing resources provided by the Supercomputing Center of the USTC.

\section{Appendix}

\subsection{Backward calculations of wave-function coefficients}
\label{backward}
For the backward calculation of $w(\mathbf{S})$, the gradient with respective to one matrix element $M_{ij}^B$ is:
\begin{equation}
	\frac{\partial \det{M^B}}{\partial M_{ij}^B}=C_{ij},
\end{equation}
where the cofactor $C_{ij}$ is an element of a $N\times N$ matrix, defined as the determinant of a matrix obtained by eliminating row i and column j from the original matrix.
Expanding the matrix $M^B$ along one column or one row with Laplace expansion, we have,
\begin{equation}
	\begin{aligned}
		\det{M^B}\cdot\delta_{ij}=\sum_{k=1}^N C_{ik}M_{jk}^B\\
		\det{M^B}\cdot\delta_{ij}=\sum_{k=1}^N C_{ki}M_{kj}^B,\\
	\end{aligned}
\end{equation}
namely,
\begin{equation}
	CM^B=M^BC=\det M^B \cdot I.
\end{equation}

Therefore, the backward of parameter $g$ is given by,
\begin{equation}
	\frac{\partial w(\mathbf{S})}{\partial g[i,k,\mathbf{s}(\mathbf{r}_i),q,\mathbf{s}(\mathbf{r}_q)]}=\mathrm{inv}(M^B)_{ik}\cdot w(\mathbf{S}),
\end{equation}
the complexity of calculating determinant and inverse of $M^B$ matrix is $\mathcal{O}(N^3)$.


\begin{thebibliography}{99}
	\bibitem{ED}
	A. L\"uscher and A. M. L\"auchli, \textit{Exact diagonalization study of the antiferromagnetic spin-1/2 Heisenberg model on the square lattice in a magnetic field}, Phys. Rev. B \textbf{79}, 195102 (2009)
	
	\bibitem{DMRG}
	L. Wang and A. W. Sandvik, \textit{Critical Level Crossings and Gapless Spin Liquid in the Square-Lattice Spin-1/2 J1-J2 Heisenberg Antiferromagnet}, Phys. Rev. Lett. \textbf{121}, 107202 (2018)

	\bibitem{QMC}
	W. M. C. Foulkes, L. Mitas, R. J. Needs, and G. Rajagopal, \textit{Quantum Monte Carlo simulations of solids}, Rev. Mod. Phys. \textbf{73}, 33 (2001)
	
	\bibitem{PEPS_review}
	J. I. Cirac, D. P.-Garc\'ia, N. Schuch, and F. Verstraete, \textit{Matrix product states and projected entangled pair states: Concepts, symmetries, theorems}, Rev. Mod. Phys. \textbf{93}, 045003 (2021)

	\bibitem{fPEPS}
	S.-J. Dong, C. Wang, Y. Han, G.-C. Guo, and L. He, \textit{Gradient optimization of fermionic projected entangled pair states on directed lattices}, Phys. Rev. B \textbf{99}, 195153 (2019)

	\bibitem{PEPS_liu}
	W.-Y. Liu, J. Hasik, S.-S. Gong, D. Poilblanc, W.-Q. Chen, and Z.-C. Gu, \textit{Emergence of Gapless Quantum Spin Liquid from Deconfined Quantum Critical Point}, Phys. Rev. X \textbf{12}, 031039 (2022)

	%\bibitem{alphago}
	%D. Silver, et. al. \textit{Mastering the game of Go with deep neural networks and tree search}, Nature, \textbf{529}, 484-489 (2016)

	\bibitem{RBM_science}
	G. Carleo and M. Troyer, \textit{Solving the quantum many-body problem with artificial neural networks}, Science \textbf{355}, 602-606 (2017)
	
	\bibitem{RBM_tomography}
	G. Torlai, G. Mazzola, J. Carrasquilla, M. Troyer, R. Melko and G. Carleo, \textit{Neural-network quantum state tomography}, Nat. Phys. \textbf{14}, 447-450 (2018)
	
	\bibitem{CNN_J1J2_1}
	X. Liang, W.-Y. Liu, P.-Z. Lin, G.-C. Guo, Y.-S. Zhang, and L. He, \textit{Solving frustrated quantum many-particle models with convolutional neural networks}, Phys. Rev. B \textbf{98}, 104426 (2018)
	
	\bibitem{CNN_J1J2_2}
	K. Choo, T. Neupert, and G. Carleo, \textit{Two-dimensional frustrated J1-J2 model studied with neural network quantum states}, Phys. Rev. B \textit{100}, 125124 (2019)
	
	\bibitem{CNN_J1J2_3}
	A. Szab\'o and C. Castelnovo, \textit{Neural network wave functions and the sign problem}, Phys. Rev. Research \textit{2}, 033075 (2020)
	
	\bibitem{CNN_J1J2_4}
	X. Liang, S.-J. Dong, and L. He, \textit{Hybrid convolutional neural network and projected entangled pair states wave functions for quantum many-particle states}, Phys. Rev. B \textit{103}, 035138 (2021)

	\bibitem{CNN_J1J2_5}
	X. Liang, M. Li, Q. Xiao, J. Chen, C. Yang, H. An and L. He, \textit{Deep learning representations for quantum many-body systems on heterogeneous hardware}, Mach. Learn.: Sci. Technol. \textbf{4}, 015035 (2023)
	
	\bibitem{RBM_J1J2}
	Y. Nomura and M. Imada, \textit{Dirac-Type Nodal Spin Liquid Revealed by Refined Quantum Many-Body Solver Using Neural-Network Wave Function, Correlation Ratio, and Level Spectroscopy}, Phys. Rev. X \textbf{11}, 031034 (2021)

	\bibitem{NN_molecules}
	K. Choo, A. Mezzacapo and G. Carleo, \textit{Fermionic neural-network states for ab-initio electronic structure}, Nat. Commun \textbf{11}, 2368 (2020)

	\bibitem{NN_solids_2}
	N. Yoshioka, W. Mizukami and F. Nori, \textit{Solving quasiparticle band spectra of real solids using neural-network quantum states}, Commun. Phys. \textbf{4} 106 (2021)

	\bibitem{DNN schrodinger 1}
	D. Pfau, J. S. Spencer, A. G. D. G. Matthews, and W. M. C. Foulkes, \textit{Ab initio solution of the many-electron Schr\"odinger equation with deep neural networks}, Phys. Rev. Research \textbf{2}, 033429 (2020)

	\bibitem{NN backflow}
	D. Luo and B. K. Clark, \textit{Backflow Transformations via Neural Networks for Quantum Many-Body Wave Functions}, Phys. Rev. Lett. \textbf{122}, 226401 (2019)

	\bibitem{DNN schrodinger 2}
	J. Hermann, Z. Sch\"atzle and F. No\'e, \textit{Deep-neural-network solution of the electronic Schr\"odinger equation}, Nat. Chem. \textbf{12}, 891-897 (2020)

	\bibitem{NN_fermi_tV}
	J. Stokes, J. R. Moreno, E. A. Pnevmatikakis, and G. Carleo, \textit{Phases of two-dimensional spinless lattice fermions with first-quantized deep neural-network quantum states}, Phys. Rev. B \textbf{102}, 205122 (2020)

	\bibitem{HFSD}
	J. R. Moreno, G. Carleo, A. Georges and J. Stokes, \textit{Fermionic wave functions from neural-network constrained hidden states}, Proc. Natl. Acad. Sci. \textbf{119} e2122059119 (2022)
	
	
	%\bibitem{autoregressive1}
	%O. Sharir, Y. Levine, N. Wies, G. Carleo and A. Shashua, \textit{Deep Autoregressive Models for the Efficient Variational Simulation of Many-Body Quantum Systems}, Phys. Rev. Lett. \textbf{124}, 020503 (2020)
	
	\bibitem{autoregressive2}
	J.-G. Liu, L. Mao, P. Zhang and L. Wang, \textit{Solving quantum statistical mechanics with variational autoregressive networks and quantum circuits}, Mach. Learn.: Sci. Technol. \textbf{2}, 025011 (2021)
	
	\bibitem{autoregressive3}
	D. Luo, Z. Chen, K. Hu, Z. Zhao, V. M. Hur and B. K. Clark, \textit{Gauge-invariant and anyonic-symmetric autoregressive neural network for quantum lattice models}, Phys. Rev. Research \textbf{5}, 013216 (2023)

	\bibitem{Hubbard_1}
	J. P. F. LeBlanc, A. E. Antipov, F. Becca, I. W. Bulk, G. K.-L. Chan, C.-M. Chung, et. al. \textit{Solutions of the Two-Dimensional Hubbard Model: Benchmarks and Results from a Wide Range of Numerical Algorithms}, Phys. Rev. X \textbf{5}, 041041 (2015)

	\bibitem{Hubbard_2}
	B.-X. Zheng, et. al. \textit{Stripe order in the underdoped region of the two-dimensional Hubbard model}, Science \textbf{358}, 1155-1160 (2017)

	\bibitem{AFQMC_1}
	M. Qin, H. Shi and S. Zhang, \textit{Coupling quantum Monte Carlo and independent-particle calculations: Self-consistent constraint for the sign problem based on the density or the density matrix}, Phys. Rev. B \textbf{94}, 235119 (2016)

	\bibitem{AFQMC_2}
	M. Qin, H. Shi, and S. Zhang, \textit{Benchmark study of the two-dimensional Hubbard model with auxiliary-field quantum Monte Carlo method}, Phys. Rev. B \textbf{94}, 085103 (2016)

	\bibitem{AFQMC_3}
	H. Xu, H. Shi, E. Vitali, M. Qin and S. Zhang, \textit{Stripes and spin-density waves in the doped two-dimensional Hubbard model: ground state phase diagram}, arXiv:2112.02187v3 (2022)

	\bibitem{backflow on position}
	Y. Kwon, D. M. Ceperley and R. M. Martin, \textit{Effects of backflow correlation in the three-dimensional electron gas: Quantum Monte Carlo study}, Phys. Rev. B \textbf{58}, 11 (1998)

	\bibitem{energy_1}
	D. A. Mazziotti, \textit{Quantum Many-Body Theory from a Solution of the N-Representability Problem}, Phys. Rev. Lett. \textbf{130}, 153001 (2023)
	
	%\bibitem{energy_2}
	%D. S. Koltun, \textit{Total Binding Energies of Nuclei, and Particle-Removal Experiments}, Phys. Rev. Lett. \textbf{28}, 182 (1972)

	\bibitem{backflow_wavefunction_1}
	L. F. Tocchio, F. Becca, A. Parola and S. Sorella, \textit{Role of backflow correlations for the nonmagnetic phase of the t-t' Hubbard model}, Phys. Rev. B \textbf{78}, 041101(R) (2008)

	\bibitem{backflow_wavefunction_2}
	L. F. Tocchio, F. Becca and C. Gros, \textit{Backflow correlations in the Hubbard model: An efficient tool for the study of the metal-insulator transition and the large-U limit}, Phys. Rev. B \textbf{83}, 195138 (2011)
	
	\bibitem{perturbation}
	J. M. Luttinger and J. C. Ward, \textit{Ground-State Energy of a Many-Fermion System. II}, Phys. Rev. \textbf{118}, 1417 (1960)

	%\bibitem{response}
	%B. Andrews and G. M\"oller, \textit{Self-similarity of spectral response functions for fractional quantum Hall states}, Proc. R. Soc. A \textbf{479}, 20230021 (2021)

	%\bibitem{conductivity1}
	%A. Mu, Z. Sun, and A. J. Millis, \textit{Optical conductivity of the two-dimensional Hubbard model: Vertex corrections, emergent Galilean invariance, and the accuracy of the single-site dynamical mean field approximation}, Phys. Rev. B \textbf{106}, 085142 (2022)

	%\bibitem{conductivity2}
	%V. Raghuraman, Y. Wang and M. Widom, \textit{An investigation of high entropy alloy conductivity using first-principles calculations}, Appl. Phys. Lett. \textbf{119}, 121903 (2021)

	%\bibitem{DMFT}
	%A. Georges, G. Kotliar, W. Krauth, and M. J. Rozenberg, \textit{Dynamical mean-field theory of strongly correlated fermion systems and the limit of infinite dimensions}, Rev. Mod. Phys. \textbf{68}, 13 (1996)

	%\bibitem{MSH}
	%P. C. Martin and J. Schwinger, \textit{Theory of Many-Particle Systems. I}, Phys. Rev. \textbf{115}, 1342 (1959)

	%\bibitem{Wick theorem}
	%R. van Leeuwen and G. Stefanucci, \textit{Wick theorem for general initial states}, Phys. Rev. B \textbf{85}, 115119 (2012)

	%\bibitem{textbook}
	%A. L. Fetter and J. D. Walecka, \textit{Quantum Theory of Many-Particle Systems}, Dover Publications, (2003)
	
	\bibitem{PEPS_GO_1}
	W.-Y. Liu, S. Dong, C. Wang, Y. Han, H. An, G.-C. Guo and L. He, \textit{Gapless spin liquid ground state of the spin-1/2 J1-J2 Heisenberg model on square lattices}, Phys. Rev. B \textbf{98}, 241109(R) (2018)

	\bibitem{PEPS_GO_2}
	W.-Y. Liu, Y.-Z. Huang, S.-S. Gong and Z.-C. Gu, \textit{Accurate simulation for finite projected entangled pair states in two dimensions}, Phys. Rev. B \textbf{103}, 235155 (2021)
	
	\bibitem{Lanczos}
	W.-J. Hu, F. Becca, A. Parola and S. Sorella, \textit{Direct evidence for a gapless Z2 spin liquid by frustrating Néel antiferromagnetism}, Phys. Rev. B \textbf{88}, 060402(R) (2013)

	\bibitem{pyscf}
	Q. Sun, et. al. \textit{Recent developments in the PySCF program package}, J. Chem. Phys. \textbf{153}, 024109 (2020)

\end{thebibliography}
\end{document}